# Investigating Effects of Common Spreadsheet Design Practices on Correctness and Maintainability

Daniel Kulesz

Sebastian Zitzelsberger

Institute of Software Technology,
University of Stuttgart, Germany

daniel.kulesz@informatik.uni-stuttgart.de

SebastianZitzelsberger@gmx.de

## ABSTRACT

*Spreadsheets are software programs which are typically created by end-users and often used for business-critical tasks. Many studies indicate that errors in spreadsheets are very common. Thus, a number of vendors offer auditing tools which promise to detect errors by checking spreadsheets against so-called Best Practices such as "Don't put constants in fomulae". Unfortunately, it is largely unknown which Best Practices have which actual effects on which spreadsheet quality aspects in which settings.*

*We have conducted a controlled experiment with 42 subjects to investigate the question whether observance of three commonly suggested Best Practices is correlated with desired positive effects regarding correctness and maintainability: "Do not put constants in formulae", "keep formula complexity low" and "refer to the left and above". The experiment was carried out in two phases which covered the creation of new and the modification of existing spreadsheets. It was evaluated using a novel construction kit for spreadsheet auditing tools called Spreadsheet Inspection Framework.*

*The experiment produced a small sample of directly comparable spreadsheets which all try to solve the same task. Our analysis of the obtained spreadsheets indicates that the correctness of "bottom-line" results is not affected by the observance of the three Best Practices. However, initially correct spreadsheets with high observance of these Best Practices tend to be the ones whose later modifications yield the most correct results.*

## 1 INTRODUCTION

Spreadsheets are software programs which are typically created by end-users with no formal training in the software engineering discipline. Nevertheless, they are often used for business-critical tasks where they can cause significant damage in case of failure [Godfrey, 1995]. Since replacing them in the near future is utopian, quality assurance is a central concern.

The use and development of spreadsheets are usually not separated and often go far beyond the creation of new spreadsheets: It is a common situation for spreadsheet users to receive spreadsheets from co-workers, to re-use old spreadsheets for new tasks or to inherit spreadsheets from people who left the company [Lawson et al., 2009]. Regardless of how a user gets in contact with a spreadsheet, the question about the spreadsheet's quality is unavoidable for any user cured from the prevalent plague of overconfidence in spreadsheets [Panko, 2003].

Many practitioners believe that "good" spreadsheets can be distinguished from "bad" ones by examining the spreadsheet's interior design against so-called Best Practices. But is observance of Best Practices really an appropriate indicator for spreadsheet quality?

We have conducted a controlled experimental study which tries to shed some light on the question to what extent the product-related spreadsheet quality aspects of correctness and maintainability are impacted by three common spreadsheet style and design rules:

- Do not put constants in formulae (also known as "isolate constants into their own cells")
- Keep formula complexity low
- Refer to the left and above ("reading direction" of references to other cells)

This paper starts with some background reflections on Best Practices for spreadsheets, measurement of spreadsheet quality and definitions for ambivalent terms. Following this, we discuss how our study is related to previous work in this area. Next, we specify the hypotheses which we tried to investigate in our study. Then, we provide a detailed description of our experiment, present our observations and discuss how far they could confirm or falsify our previous hypotheses. Finally, we show threats to validity and summarize the key contributions of this work.

## 2 BACKGROUND

### 2.1 Best Practice Uncertainty

Unlike for professional software, there are no generally accepted "Best Practices" or design patterns in the spreadsheet world. While there is a large number of publications which proclaim Best Practices for spreadsheet design such as 'Don't put constants in formulae', 'Keep formula complexity low' or 'Avoid named ranges' [Raffensberger 2001] [O'Beirne, 2005] [Bovey et. al., 2009] [Powell and Baker, 2010], many of the proposed practices contradict each other and there are discussions going on about whether "universal" Best Practices exist at all [Colver, 2004]. Our previous paper [Kulesz, 2011] describes this issue in more detail.

When comparing Best Practices for spreadsheets with Design Patterns for professional Software [Gamma et. al., 1994], one major difference becomes apparent: Design Patterns make statements about their (positive) intent, their applicability and related (negative) consequences while Best Practices often name only positive influences.

Apart from the relevance of the qualitative impact of Best Practices for spreadsheets in general, there is also a more fine-grained question about calibration of such rules: For instance, there might be exceptions for "credible constants" such as "1". But which constants are really credible? Similarly, when checking complexity metrics [Bregar, 2004] such as the number of operands in formulae or the nesting level of expressions therein to measure formula complexity, feasible thresholds are required.

Despite these uncertainties, a number of commercial vendors offer static analysis tools for spreadsheets such as [SpreadsheetInnovations, 2012] or [Codematic, 2012]. Most of those tools check spreadsheets against such rules as previously described. But both the rules and the thresholds are hardwired into the tools. As a result, the number of reported rule violations can vary greatly between different tools even for rules which seem to be the same [Zitzelsberger, 2012].

The practical consequence for end-users is that they have to trust tool vendors not only to have implemented the rules correctly but also to have selected appropriate, effective rules and calibrated them sensibly. But it is highly questionable that this trust can be satisfied when there are very few scientific studies which confirm the effectiveness of such rules or deal with the investigation of sensible calibration thresholds.

## 2.2 Evaluating Spreadsheet Quality and Errors

With the exception of criminals with fraudulent intent, errors in spreadsheets are not desired by users. But what exactly are spreadsheet errors? Over the past years, there have been many efforts to create taxonomies for classifying different types of errors in spreadsheets. Powell, Baker and Lawson [Powell et al., 2008] stated in 2008 that there were many classifications for spreadsheet errors back then but no classification was widely accepted.

This situation has not changed significantly since then. Nevertheless, the spreadsheet error classification quoted most frequently seems to be the "Revised Panko and Halverson Taxonomy of Spreadsheet Errors" [Panko and Aurigemma, 2010] and much derived work like [Przasnyski et. al., 2011] try to extend it. One major characteristic of this taxonomy is the distinction between "quantitative errors" which manifest themselves in wrong "bottom-line" results and "qualitative errors" which have only indirect negative effects like "making the model more prone to misinterpretation" or "making the model more difficult to update" [Rajalingham, 2005]. There are numerous publications which stress the importance of these "qualitative" errors [Panko, 1988] [Pryor, 2003] [Correia and Ferreira, 2011].

We do not doubt that having a widely accepted taxonomy for the classification of spreadsheet errors is essential. But since errors are something *negative*, discussing error taxonomies requires that there is agreement about which *positive* quality characteristics are desired.

## 2.3 Definitions

There are different approaches towards defining quality in general [Garvin, 1984]. We refer to the definition of quality according to the DIN 55350 standard [DIN, 2008]: "Quality is the sum of properties and characteristics of a product or an activity which contribute to the suitability of satisfying described needs" (translated from German). We use the term *quality aspect* for a single characteristic of a product or activity.

When talking about deviations from desired quality aspects, often the terms "error", failure", "fault" and "defect" can be encountered. We use these terms as they are defined in IEEE Std 1044™-2009:

**error**: A human action that produces an incorrect result.

**failure**: (A) Termination of the ability of a product to perform a required function or its inability to perform within previously specified limits. (B) An event in which a system or system component does not perform a required function within specified limits.

**fault**: A manifestation of an error in software.

**defect**: An imperfection or deficiency in a work product where that work product does not meet its requirements or specifications and needs to be either repaired or replaced.

Therefore, even if we find incorrect "bottom-line" results in spreadsheets, we cannot make statements about errors if we only analyze spreadsheets in their post-creation state without examining the human actions which caused them.

Our work focuses on two quality aspects: correctness and maintainability. We define them as:

**correctness**: The degree to which the "bottom-line" results computed by a spreadsheet conform to manually computed results for given test scenarios.

**maintainability**: The sum of internal factors of a spreadsheet which contribute to its chance of maintaining correctness throughout modifications.

# 3 RELATED WORK

A general problem for the evaluation of spreadsheet quality is the low public availability of comparable spreadsheets with known errors. The often cited EUSES Spreadsheet corpus [Fisher II and Rothermel, 2005] is not suitable because it provides a huge collection of random spreadsheets found on the Internet for which no specifications are available. Therefore, it is hardly possible to make judgements about the correctness of its spreadsheets. Panko provides two corpuses from exercises with students, the so-called "Galumpke corpus" with 82 spreadsheets and the "wall corpus" with 150 spreadsheets [Panko, 2000]. Teo and Tan reproduced a closely adapted version of the "wall corpus" with their students that comprises of 168 spreadsheets [Teo and Tan, 1997]. The task for "Galumpke" spreadsheets demanded considerable knowledge about the accounting domain from the subjects, while the "wall" task was much more domain-independent. Nevertheless, both tasks used a very short textual description of the problem and involved almost no data. This limits their comparability with many real-world spreadsheets.

There are only few published studies which examine the actual effects of single Best Practices on single spreadsheet quality aspects:

- Teo and Tan [Teo and Tan, 1997] studied the effects of hardcoded constants which were already mentioned as being an issue by Panko in 1988 [Panko, 1998]. Their subjects first performed an exercise very similar to Panko's wall problem [Panko, 1996] followed by second task with a "what if"-analysis. One key finding of their study was that there are significant positive correlations between hardcoded constants in formulae and logic errors.

- In another experiment Panko [Panko, 1999] observed that errors in more complex formulae had a higher chance for staying undetected after inspections than errors in simpler formulae.

- McKeever and McDaid investigated some effects of range naming conventions on reliability, development time and debugging performance [McKeever et. al., 2009] [McKeever and McDaid, 2010] [McKeever and McDaid, 2011].

- Kruck [Kruck, 2006] provides partial evidence for her hypothesis that high formula complexity has a negative effect on correctness (she refers to our notion of correctness using the term "accuracy").

Because only few effects have been investigated to date, there are various views about which Best Practice violations should be regarded only as "stylistic" problems or as "common sources of error". This becomes particularly evident in Blayney's comparison of the perception among researchers about the "Hard coding constants" rule [Blayney, 2006].

Also, numerous publications proclaim that "context is important". But we could not find any studies which evaluated actual Best Practice effectiveness for different types of spreadsheets or domain contexts. Also, we could not find any publications which deal with the effects of different calibrations of threshold variables commonly used in tools which check for compliance with Best Practices.

## 4 RESEARCH QUESTIONS

We designed our study with the goal of contributing to the following main research question:

- RQ1: Can observance of spreadsheet Best Practices serve as a dependable quality indicator?

Since the term quality can refer to a large number of particular quality aspects, we have limited the study to the following two more fine-grained questions:

- RQ2: Does observance of certain Best Practices affect correctness?
- RQ3: Does observance of certain Best Practices affect maintainability?

Of course, it would require a tremendous amount of work to evaluate these research questions for *all* Best Practices suggested in literature. Therefore, we restricted our study to the three commonly suggested Best Practices mentioned in the introduction ("Do not put constants in formulae", "Keep formula complexity low" and "Refer to the left and above").

## 5 EXPERIMENT

This section describes an experiment which we conducted to approach the research questions. Due to size limitations, we can only describe the most relevant facts in this section. The full task descriptions, supplemental materials and resulting spreadsheets are available from our website. [ISTE, 2012]

### 5.1 Overview

We conducted an experiment where a total of 42 subjects performed a number of tasks by means of a spreadsheet software. The experiment was performed with one subject at a time. Each subject received a short personal briefing. The whole study took 7 months between September 2011 and March 2012. Actually there were two different experiments to which we refer as "phase 1" and "phase 2", because experiments in phase 2 built upon artifacts from phase 1. Another major difference is that phase 2 was controlled much more strictly in terms of environmental variables such as available time for processing, physical location, spreadsheet software, and screen resolution.

Both phases were themed around a typical "which product to buy" scenario in the context of passenger cars. The subjects were given a written task description on a single A4 paper sheet together with 3 (phase 1) or 2 (phase 2) printed excerpts from test reports, where popular passenger cars were tested and given grades in so called "testimonials". The grades were given in the German grading system (1.0 is best, 6.0 is worst) for 7 different categories like "comfort" or "handling". Each category had 2 to 6 single disciplines, i.e. the grade for comfort consisted of "suspension", "seats", "interior noise" and "heating / air conditioning" which were all given individual grades. Also, there were two "major categories": The first one consisted of 7 categories with their individual grades while the second one had 5 individual grades without having a category.

### 5.2 Phase 1

18 subjects took part in phase 1. They were given the task to build a new spreadsheet from scratch which should answer the "which car to buy" question based on user-definable custom weights for all grades found in the "testimonial". For instance, one user of the spreadsheet might be interested in a very safe car, giving this category a high weight while another user might care much more about engine performance; the job of the spreadsheet was to compute "personal grades", taking into account the weights and the grades of the testimonial. Furthermore, the following additional requirements were stated in the task description:

- The spreadsheet shall support the comparison between the three cars given but it should be designed with extendibility in mind to support additional cars in the future
- The spreadsheet may only use formulae (VBA macros are forbidden)
- All computed (sub-)grades must be accurate to the first decimal
- The spreadsheet should be clearly arranged and visually appealing

We didn't strictly control environmental variables in phase 1 because we believe that spreadsheets are created in a variety of totally different environments today and we wanted to reflect this diversity in our spreadsheet samples. Instead, we conducted the experiments in a LibreOffice 3.3 environment running in a "live" Linux environment which was booted from a USB pen drive. This allowed us to run the experiment at different locations like homes or office workplaces and not just in our university. This way, we had a vast variety of computer screens, ranging from 14" XGA (1024x768) laptop screens to 24" WUXGA (1920x1200) professional monitors.

The subjects were given a *rough* time limit of one hour. We did not interrupt subjects who took longer though. Most subjects kept to this time frame with very few exceptions: One subject took almost two hours to finish, while another was done in just 25 minutes.

## 5.3 Phase 2

In phase 2 we had a total of 28 subjects who accepted the challenge to modify an existing spreadsheet from phase 1. We used only four different phase 1 spreadsheets as "input sheets" for phase 2 though (see Figure 2), so each input sheet was modified by 7 different subjects, resulting in a total of 28 modified spreadsheets. No subject modified more than one input sheet to avoid learning effects.

The input sheets were not selected arbitrarily: We tried to find mostly correct candidates which solved the task but did so in fundamentally different ways regarding their visual and interior (formulae) design.

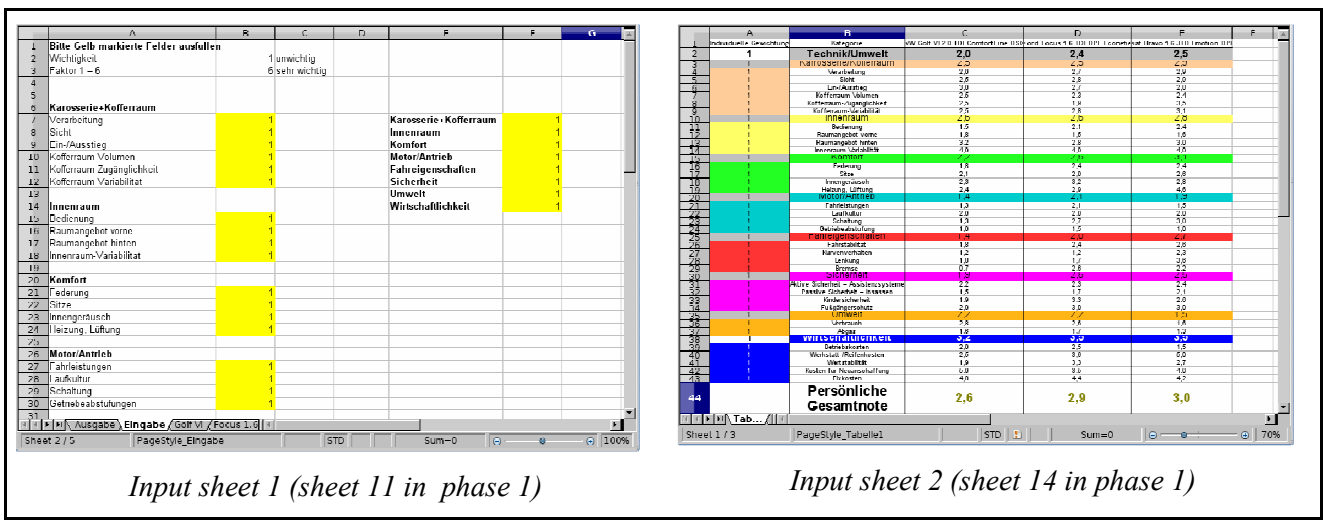

*Input sheet 1 (sheet 11 in phase 1)*          *Input sheet 2 (sheet 14 in phase 1)*

*Input sheet 3 (sheet 2 in phase 1)*      *Input sheet 4 (sheet 4 in phase 1)*

*Figure 2: The selected "input sheets" from phase 1*

The task of Phase 2 subjects was again to solve a number of tasks based on a written description on an A4 paper sheet, accompanied by additional "car testimonials". The tasks were:

- Extend the car comparison by the two additional cars.
- Introduce a new variable "annual mileage" which creates a "handicap" multiplier for two of the grades. (We provided a table of different ranges for annual mileages and appropiate multipliers.)
- Change the weights so that car 1 becomes the winner and car 3 gets the lowest overall recommendation.
- Distinguish the winner car visually from the others by increasing the font size and applying a different font color.

This time, the subjects were given a strict time limit of 55 minutes. The environment used was Excel 2003 running on Windows XP with a 19" SXGA (1280x1024) screen in our computer lab.

### 5.4 Selection of Subjects

"You know a bit how to use Excel, don't you?" - this was the introductory question we asked people to invite them to join our experiment. We mostly asked people we knew personally and who lived in Stuttgart. This way we got a mixed population of engineers, secretaries, students, teachers, artists and various other subjects with ages ranging from 20 to 55. Of course, we attracted many people from our "natural" environment (a computer science faculty) this way, but we put increased effort into keeping the percentage of "IT people" low.

After they finished the experiment, all subjects were asked to fill out a questionnaire about their use of spreadsheets and their previous knowledge. The questionnaire was roughly based on the questionnaire used in the study of [Lawson et. al., 2009] but we reduced the number of questions and skipped some of the topics discussed there to make it much more compact (our version has a length of just 4 pages). By using the questionnaire we tried to balance the distribution of existing experience among the subjects, but we were only partly successful as the analysis of two key questions provided in Figure 3 shows.

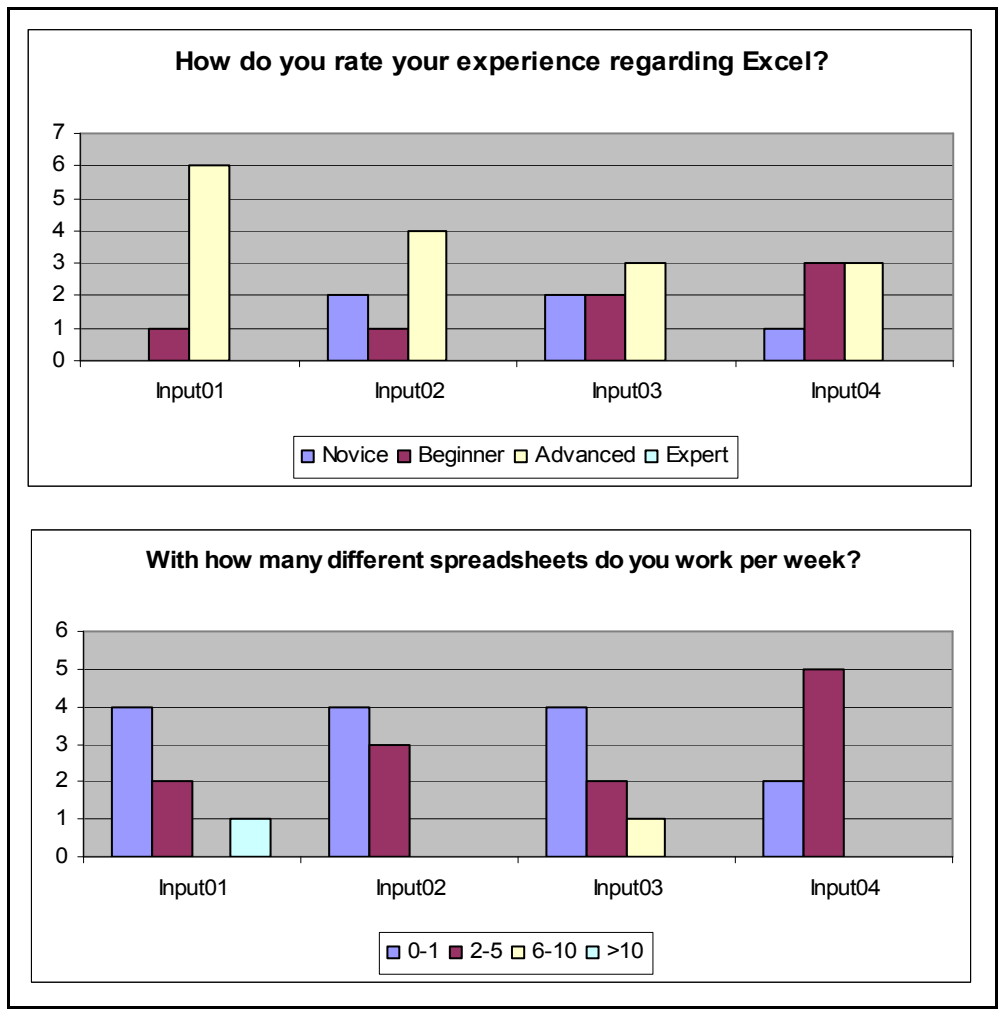

*Figure 3: Answer distribution for 2 of the 27 questions asked in our questionnaire*

Subjects who took part in phase 1 experiments were not reused in phase 2 to avoid learning effects. We made one exception though: The authors of the selected "input sheets" were asked to participate in phase 2 after a break of at least two weeks. We decided to reuse them because we were interested in performance differences of the original authors who were supposed to know the input sheets better than the people who saw the input sheets for the first time. But unless stated otherwise, we excluded the performance of the original authors from our observations and illustrations for Phase 2 (this is also true for Figure 3).

## 6 RESULTS

### 6.1 Analysis Method

In order to verify the correctness of the spreadsheets we obtained from the experiments, we defined one test scenario for phase 1 and four test scenarios for phase 2. The test scenarios were pretty rigid and included testing for boundary values. Each test scenario provided input values for all the "weight variables" (in both phases) and the "annual mileage" (only for phase 2). We calculated the expected results for the "personal grades" in all the scenarios manually (without a spreadsheet). Of course, the spreadsheets could possibly show more errors when more test cases were exectued.

We entered the input values as defined in the test scenarios and recorded the actual results obtained for the computed "personal recommendation" grades for each car. Finally, we made a comparison table which summarizes all values for all cars in all scenarios from all experiments and visualized the deviations. We also inspected the grades which had to be entered manually by the subjects for data errors.

While this analysis was very clear for phase 1, we had to make some exceptions for phase 2 because, as it turned out, we had selected two slightly flawed "input sheets" for phase 2: One input sheet contained a data entry error and the other one did not distinguish between categories and "major categories". Because it was unrealistic to expect the phase 2 participants to detect these problems (they would have needed the data sheets), we used different "expected values" for the modified spreadsheets which were derived from those flawed input sheets. These "expected values" were calculated under the assumption that all input sheets were correct, including the flawed ones.

Please note that we did not analyse other aspects of correctness apart from the personal grades. For instance, we could also have checked whether the "winner car" was always highlighted or whether the ranking of the cars was in the right order.

Apart from these manual correctness measures, we also collected two objective metrics of the spreadsheets: Number of non-empty cells and number of formulae. Finally, we counted the violations of Best Practice rules and put them in relation to the objective metrics. All analyses were done automatically using the tool described in the next section.

## 6.2 Tools and Rules

As already explained in the background sections, the formulation of checking rules for Best Practices is not clear and unambiguous because of the threshold variables involved and general questions about suitable metrics, i.e. for measuring formula complexity.

We used an application named "Example Testing Center" (ETC) which is built upon a novel spreadsheet analysis framework named "Spreadsheet Inspection Framework" (SIF). Both the application and the framework were developed in a diploma thesis by Sebastian Zitzelsberger [Zitzelsberger, 2012]. SIF differs from existing spreadsheet analysis tools in a few ways:

- It is open source and freely available from GitHub [SIF, 2012].
- Internally it uses an independent spreadsheet model specially designed for quality analysis of spreadsheets. Adapters for common spreadsheet file formats like XLS or XLSX are provided.
- The thresholds used in its rules are customizable by the user.
- It uses the metaphor of "MOT garage" and tries to transfer core concepts used for car general inspections to spreadsheet inspections. It is meant as a toolkit for such "spreadsheet inspection centers". ETC is just a demonstration example of such a center.
- SIF contains an option for reporting repeated violations (i.e. a formula with a constant was copied over to 20 neighbor cells) by arranging them in so-called "violation groups".
- Additional rules can be easily implemented by using one of SIF's several extension points.

A full description of this work is provided in [Zitzelsberger, 2012]. For understanding the evaluation of our experiments, it is important to know how the checking rules for the Best Practices stated in the research questions are implemented in the current version of ETC and SIF:

**Do not put constants in formulae**:

SIF inspects each formula and looks for constants in it. It reports all constants for configuration 1 as defects but ignores the constants "1" and all constants used in the "INDEX" function for configuration 2.

**Keep formula complexity low**:

To measure formula complexity, SIF counts the number of operations (operands and functions) and the maximum nesting level of operations. In configuration 1, formulae exceeding 5 operations are counted as defects while in configuration 2 only 2 operations are allowed per formula. For both configurations the maximum allowed nesting level is 2.

**Refer to the left and above**:

All formulae are inspected by SIF and in case SIF finds references to other cells, it reports all referenced cells which are located right or below the referencing cell as defects. Similarly, all referenced cells in other worksheets ("tabs") to the right of the worksheet with the formula are reported as defects. Optionally one of these checks can be disabled but for configuration 1 and 2 we left both checks enabled.

## 6.3 Observations

Figure 4 shows the results we obtained after the analysis of the phase 1 experiments. The results obtained from the phase 2 experiments are shown in Figure 5. Spreadsheets which do not satisfy basic functional requirements (i.e. offer no possibility to specify the weights or an annual mileage) are marked with an "X".

As expected, changes to the thresholds of the used Best Practices did change the absolute numbers of detected violations but this does not make a notable difference for the conclusions which can be drawn. Therefore, all the following observations refer to the results obtained by using configuration 1.

Below are some general remarks about the results of both phases:

- In phase 1, 71% of the subjects managed to produce a functional and feature-complete spreadsheet. In phase 2 it were even 79%.
- Data entry errors were very rare: Only 12% of the sheets in phase 1 and 17% of the sheets in phase 2 contained data entry errors. No spreadsheet had more than one data error although the subjects had to enter a total of 99 data values in phase 1 and 66 data values in phase 2 which corresponds to a cell error rate (CER) of 1 to 1.6 %. This is consistent with other studies [Powell et. al., 2008].
- The spreadsheets in phase 1 had a median of 397 non-empty cells. The biggest spreadsheet had 780 non-empty cells while the smallest spreadsheet had 214. This number increased after the modifications in phase 2 by a median of 53%. In the median, these increases did not differ remarkably between modifications across the different input sheets.
- The spreadsheets in phase 1 produced a median of 69 for cells containing formulae. The largest spreadsheet had 277 cells with formulae while the smallest spreadsheet had just 3. This number increased after the modifications in phase 2 by a median of 65%. There were noteworthy differences between the different input sheets: In the median the modifications of input sheet 4 grew by 52%, while the modifications of input sheet 2 grew by 77%.
- 25% of the subjects in phase 2 either fully observed the "no constants in formulae" rule for all their formulae or ignored it everywhere.
- Defect rates of modified spreadsheets are strongly influenced by the defect rates of their input sheets.

- Although we accidentally exercised input sheet 1 with far more subjects who claimed to have advanced spreadsheet skills (when compared to the other input sheets where the population was more balanced), all of the solutions for input sheet 1 failed in all test scenarios.

|  | 1 | 2 | 4 | 5 | 6 | 7 | 8 | 9 | 10 | 11 | 12 | 13 | 14 | 15 | 16 | 17 | 18 |
|---|---|---|---|---|---|---|---|---|---|---|---|---|---|---|---|---|---|
| # of data errors | 0 | 0 | 0 | 0 | 0 | 0 | 0 | 0 | 0 | 0 | 1 | 0 | 0 | 0 | 0 | 1 | 0 |
| # of cells | 220 | 530 | 397 | 592 | 222 | 497 | 214 | 780 | 223 | 741 | 507 | 382 | 220 | 336 | 237 | 444 | 419 |
| # of formulae | 3 | 163 | 166 | 184 | 30 | 126 | 27 | 277 | 3 | 241 | 69 | 7 | 30 | 54 | 7 | 223 | 87 |
| Failed in # of scenarios | 1 | 0 | 0 | 1 | 1 | X | X | 1 | X | 1 | X | 1 | 0 | 0 | 1 | 1 | X |

**Configuration 1**

|  | 1 | 2 | 4 | 5 | 6 | 7 | 8 | 9 | 10 | 11 | 12 | 13 | 14 | 15 | 16 | 17 | 18 |
|---|---|---|---|---|---|---|---|---|---|---|---|---|---|---|---|---|---|
| Formula Complexity | 0 | 1 | 27 | 30 | 0 | 102 | 27 | 3 | 0 | 27 | 0 | 3 | 30 | 54 | 0 | 0 | 3 |
| .. relative to # of formulae | 0% | 1% | 16% | 16% | 0% | 81% | 100% | 1% | 0% | 11% | 0% | 43% | 100% | 100% | 0% | 0% | 3% |
| No Constants In Formulae | 0 | 10 | 67 | 0 | 0 | 126 | 27 | 0 | 3 | 0 | 3 | 7 | 30 | 0 | 3 | 0 | 3 |
| .. relative to # of formulae | 0% | 6% | 40% | 0% | 0% | 100% | 100% | 0% | 100% | 0% | 4% | 100% | 100% | 0% | 43% | 0% | 3% |
| Reading direction | 0 | 27 | 27 | 184 | 27 | 27 | 27 | 27 | 0 | 198 | 1 | 4 | 27 | 54 | 0 | 47 | 0 |
| .. relative to # of formulae | 0% | 17% | 16% | 100% | 90% | 21% | 100% | 10% | 0% | 82% | 1% | 57% | 90% | 100% | 0% | 21% | 0% |

**Configuration 2**

|  | 1 | 2 | 4 | 5 | 6 | 7 | 8 | 9 | 10 | 11 | 12 | 13 | 14 | 15 | 16 | 17 | 18 |
|---|---|---|---|---|---|---|---|---|---|---|---|---|---|---|---|---|---|
| Formula Complexity | 3 | 4 | 67 | 30 | 2 | 102 | 27 | 27 | 0 | 27 | 0 | 3 | 30 | 54 | 0 | 3 | 3 |
| .. relative to # of formulae | 100% | 2% | 40% | 16% | 7% | 81% | 100% | 10% | 0% | 11% | 0% | 43% | 100% | 100% | 0% | 1% | 3% |
| No Constants In Formulae | 0 | 0 | 67 | 0 | 0 | 126 | 0 | 0 | 3 | 0 | 3 | 7 | 0 | 0 | 3 | 0 | 3 |
| .. relative to # of formulae | 0% | 0% | 40% | 0% | 0% | 100% | 0% | 0% | 100% | 0% | 4% | 100% | 0% | 0% | 43% | 0% | 3% |
| Reading direction | 0 | 27 | 27 | 184 | 27 | 27 | 27 | 27 | 0 | 198 | 1 | 4 | 27 | 54 | 0 | 47 | 0 |
| .. relative to # of formulae | 0% | 17% | 16% | 100% | 90% | 21% | 100% | 10% | 0% | 82% | 1% | 57% | 90% | 100% | 0% | 21% | 0% |

*Figure 4: Analysis of the phase 1 spreadsheets*

Regarding correctness, the following conclusions can be drawn:

- Out of the 12 feature-complete spreadsheets in phase 1 only 4 (33%) produced a correct result in the single scenario evaluated. We obtained a similar value for phase 2 where only 4 out of 19 feature-complete spreadsheets (21%) were correct in all three evaluation scenarios. Another 3 spreadsheets (16%) only failed in one scenario, one (5%) failed in two scenarios and the remaining ones (58%) failed in all three scenarios. It is questionable though to which degree such partial failures can be considered useful spreadsheets because it completely depends on the later usage whether or not the errors contained in them would cause an impact.
- Correct spreadsheets cannot be clearly identified by inspecting the percentage of complex fomulae: For correct spreadsheets in both phases the complexity varies between 0 and 100%. Vice versa, faulty spreadsheets cannot be clearly identified by having a high percentage of complex formulae.
- Spreadsheets containing formulae with constants are not a dependable indicator: The percentage of formulae with constants varies between 0 and 100% for correct spreadsheets in phase 1 and between 3 to 100% for correct spreadsheets in phase 2. Vice versa, faulty spreadsheets do not have a high percentage of constants in formulae.
- The observations of formula complexity and constants in formulae can be mostly transfered to "reading direction" as well.
- Only a single spreadsheet in phase 1 showed no defects for all evaluted Best Practices and only 28% of the spreadsheets in this phase showed less than 10 defects. Interestingly, neither one of these spreadsheets produces correct results.

|  | input01 | | | | | | | input02 | | | | | | |
|---|---|---|---|---|---|---|---|---|---|---|---|---|---|---|
| subject | 1 | 2 | 3 | 4 | 5 | 6 | 21 | 7 | 8 | 10 | 11 | 13 | 19 | 20 |
| # of data errors | 0 | 0 | 1 | 0 | 0 | 0 | 1 | 0 | 1 | 0 | 0 | 0 | 0 | 0 |
| # of cells | 66581 | 1111 | 1134 | 1417 | 1140 | 1077 | 1116 | 1643 | 354 | 338 | 308 | 322 | 315 | 332 |
| relative cell increase | 8885% | 50% | 53% | 91% | 54% | 45% | 51% | 647% | 61% | 54% | 40% | 46% | 43% | 51% |
| # of formulae | 283 | 405 | 407 | 465 | 410 | 354 | 409 | 53 | 32 | 67 | 50 | 52 | 53 | 60 |
| relative formulae increase | 17% | 68% | 69% | 93% | 70% | 47% | 70% | 77% | 7% | 123% | 67% | 73% | 77% | 100% |
| failed in # of scenarios | X | 3 | 3 | 1 | 3 | 3 | 3 | 1 | X | 1 | X | 3 | 3 | 0 |
| # of wrong results / scenario | X,X,X | 1,2,1 | 1,1,1 | 2,4,2 | 2,3,2 | 2,4,1 | 2,4,2 | 0,5,0 | X,X,X | 0,5,0 | X,X,X | 4,4,4 | 1,1,4 | 0,0,0 |

**Configuration 1**

|  | 1 | 2 | 3 | 4 | 5 | 6 | 21 | 7 | 8 | 10 | 11 | 13 | 19 | 20 |
|---|---|---|---|---|---|---|---|---|---|---|---|---|---|---|
| Formula Complexity | 45 | 45 | 45 | 47 | 47 | 45 | 45 | 52 | 32 | 52 | 50 | 52 | 52 | 60 |
| .. relative to # of formulae | 16% | 11% | 11% | 10% | 11% | 16% | 11% | 98% | 100% | 78% | 100% | 100% | 98% | 100% |
| relative defct increase | 67% | 67% | 67% | 74% | 74% | 104% | 67% | 73% | 7% | 73% | 67% | 73% | 73% | 100% |
| No Constants In Formulae | 1 | 2 | 0 | 2 | 5 | 10 | 10 | 52 | 32 | 52 | 50 | 52 | 2 | 60 |
| .. relative to # of formulae | 0% | 0% | 0% | 0% | 1% | 3% | 2% | 98% | 100% | 78% | 100% | 100% | 4% | 100% |
| relative defct increase | ~ | ~ | ~ | ~ | ~ | ~ | ~ | 73% | 7% | 73% | 67% | 73% | -93% | 100% |
| Reading direction | 231 | 330 | 330 | 380 | 332 | 290 | 340 | 45 | 27 | 55 | 45 | 45 | 45 | 45 |
| .. relative to # of formulae | 82% | 81% | 81% | 82% | 81% | 82% | 83% | 85% | 84% | 82% | 90% | 87% | 85% | 75% |
| relative defct increase | 17% | 67% | 67% | 92% | 68% | 46% | 72% | 67% | 0% | 104% | 67% | 67% | 67% | 67% |

**Configuration 2**

|  | 1 | 2 | 3 | 4 | 5 | 6 | 21 | 7 | 8 | 10 | 11 | 13 | 19 | 20 |
|---|---|---|---|---|---|---|---|---|---|---|---|---|---|---|
| Formula Complexity | 45 | 45 | 45 | 47 | 50 | 55 | 45 | 52 | 32 | 52 | 50 | 52 | 52 | 60 |
| .. relative to # of formulae | 16% | 11% | 11% | 10% | 12% | 16% | 11% | 98% | 100% | 78% | 100% | 100% | 98% | 100% |
| relative defct increase | 67% | 67% | 67% | 74% | 85% | 104% | 67% | 73% | 7% | 73% | 67% | 73% | 73% | 100% |
| No Constants In Formulae | 0 | 2 | 0 | 2 | 5 | 10 | 10 | 2 | 0 | 2 | 0 | 2 | 2 | 10 |
| .. relative to # of formulae | 0% | 0% | 0% | 0% | 1% | 3% | 2% | 4% | 0% | 3% | 0% | 4% | 4% | 17% |
| relative defct increase | ~ | ~ | ~ | ~ | ~ | ~ | ~ | ~ | ~ | ~ | ~ | ~ | ~ | ~ |
| Reading direction | 231 | 330 | 330 | 380 | 332 | 290 | 340 | 45 | 27 | 55 | 45 | 45 | 45 | 45 |
| .. relative to # of formulae | 82% | 81% | 81% | 82% | 81% | 82% | 83% | 85% | 84% | 82% | 90% | 87% | 85% | 75% |
| relative defct increase | 17% | 67% | 67% | 92% | 68% | 46% | 72% | 67% | 0% | 104% | 67% | 67% | 67% | 67% |

|  | input03 | | | | | | | input04 | | | | | | |
|---|---|---|---|---|---|---|---|---|---|---|---|---|---|---|
| subject | 12 | 14 | 15 | 16 | 17 | 18 | 22 | 23 | 24 | 25 | 26 | 27 | 28 | 29 |
| # of data errors | 0 | 0 | 0 | 0 | 0 | 0 | 0 | 0 | 0 | 0 | 0 | 0 | 0 | 0 |
| # of cells | 664 | 835 | 812 | 1002 | 842 | 850 | 904 | 570 | 595 | 576 | 598 | 604 | 616 | 700 |
| relative cell increase | 25% | 58% | 53% | 89% | 59% | 60% | 71% | 44% | 50% | 45% | 51% | 52% | 55% | 76% |
| # of formulae | 169 | 270 | 265 | 285 | 302 | 267 | 292 | 251 | 269 | 252 | 255 | 252 | 253 | 187 |
| relative formulae increase | 46% | 66% | 63% | 75% | 85% | 64% | 79% | 51% | 62% | 52% | 54% | 52% | 52% | 13% |
| failed in # of scenarios | 3 | 0 | 0 | X | 2 | 1 | 0 | 0 | 3 | 3 | 3 | 0 | 3 | X |
| # of wrong results / scenario | 3,4,4, | 0,0,0 | 0,0,0 | X,X,X | 0,5,1 | 0,0,1 | 0,0,0 | 0,0,0 | 4,4,4 | 3,5,4 | 5,5,5 | 0,0,0 | 5,5,5 | X,X,X |

**Configuration 1**

|  | 12 | 14 | 15 | 16 | 17 | 18 | 22 | 23 | 24 | 25 | 26 | 27 | 28 | 29 |
|---|---|---|---|---|---|---|---|---|---|---|---|---|---|---|
| Formula Complexity | 1 | 1 | 3 | 1 | 3 | 3 | 13 | 55 | 56 | 52 | 50 | 47 | 47 | 27 |
| .. relative to # of formulae | 1% | 0% | 1% | 0% | 1% | 1% | 4% | 22% | 21% | 21% | 20% | 19% | 19% | 14% |
| relative defct increase | 0% | 0% | 200% | 0% | 200% | 200% | 1200% | 104% | 107% | 93% | 85% | 74% | 74% | 0% |
| No Constants In Formulae | 16 | 22 | 9 | 10 | 12 | 12 | 14 | 96 | 102 | 87 | 90 | 87 | 87 | 88 |
| .. relative to # of formulae | 9% | 8% | 3% | 4% | 4% | 4% | 5% | 38% | 38% | 35% | 35% | 35% | 34% | 47% |
| relative defct increase | 60% | 120% | -10% | 0% | 20% | 20% | 40% | 43% | 52% | 30% | 34% | 30% | 30% | 31% |
| Reading direction | 35 | 45 | 60 | 45 | 60 | 45 | 71 | 55 | 59 | 60 | 45 | 55 | 47 | 27 |
| .. relative to # of formulae | 21% | 17% | 23% | 16% | 20% | 17% | 24% | 22% | 22% | 24% | 18% | 22% | 19% | 14% |
| relative defct increase | 30% | 67% | 122% | 67% | 122% | 67% | 163% | 104% | 119% | 122% | 67% | 104% | 74% | 0% |

**Configuration 2**

|  | 12 | 14 | 15 | 16 | 17 | 18 | 22 | 23 | 24 | 25 | 26 | 27 | 28 | 29 |
|---|---|---|---|---|---|---|---|---|---|---|---|---|---|---|
| Formula Complexity | 6 | 16 | 13 | 6 | 8 | 8 | 18 | 96 | 92 | 87 | 90 | 87 | 87 | 67 |
| .. relative to # of formulae | 4% | 6% | 5% | 2% | 3% | 3% | 6% | 38% | 34% | 35% | 35% | 35% | 34% | 36% |
| relative defct increase | 50% | 300% | 225% | 50% | 100% | 100% | 350% | 43% | 37% | 30% | 34% | 30% | 30% | 31% |
| No Constants In Formulae | 6 | 12 | 7 | 0 | 2 | 2 | 0 | 96 | 102 | 87 | 90 | 87 | 87 | 88 |
| .. relative to # of formulae | 4% | 4% | 3% | 0% | 1% | 1% | 0% | 38% | 38% | 35% | 35% | 35% | 34% | 47% |
| relative defct increase | ~ | ~ | ~ | ~ | ~ | ~ | ~ | 43% | 52% | 30% | 34% | 30% | 30% | 31% |
| Reading direction | 35 | 45 | 60 | 45 | 60 | 45 | 71 | 55 | 59 | 60 | 45 | 55 | 47 | 27 |
| .. relative to # of formulae | 21% | 17% | 23% | 16% | 20% | 17% | 24% | 22% | 22% | 24% | 18% | 22% | 19% | 14% |
| relative defct increase | 30% | 67% | 122% | 67% | 122% | 67% | 163% | 104% | 119% | 122% | 67% | 104% | 74% | 0% |

*Figure 5: Analysis of the phase 2 spreadsheets (phase 1 authors are marked in light blue)*

Regarding maintainability, the following conclusions can be drawn:

- There is a trend that input sheets which conform to Best Practices have a higher chance of maintenance success than input sheets which do not: Input sheet 3 had the lowest number of defects (1%, 6%, 17%) and the most correct modifications (2/6). Input sheet 4 had 1/6 correct modifications and a low number of defects (16%, 40%, 16%). Input sheet 2 had only one successful and two "partially successful" modifications while having an enormous defect rate (100%, 100%, 90%). Input sheet 1 is an exception: It did not even have a single partially successful modification although its defect rate is moderate (11%, 0%, 82%).
- Maintenance success does not seem to be influenced by defect increase: Defects increase at completely varying rates in both successful and unsuccessful modifications.
- The original authors of the phase 2 input sheets did not perform much better than the other subjects who were unfamiliar with both the task and the particular spreadsheets. Although two authors managed to produce correct results for all scenarios, they also introduced about the same amount of new defects as the other subjects.

In summary, our results indicate that there seems to be no connection between the observance of the Best Practices selected for this study (at least in the way they were interpreted) and the correctness of the "bottom-line" results. We tried several variations for all the incorporated thresholds but were not able to find a set of thresholds which would clearly distinguish the "good" spreadsheets from the "bad" ones.

Regarding maintainability, there is a visible trend that modifications of spreadsheets which observe the studied Best Practices have higher chances to be successful. This is especially promising as relative defect rates stay mostly the same after modification.

**6.4 Threats to Validity**

We are well aware of the fact that the results from this study are not very dependable due to the following shortcomings which would require further efforts in future work:

- Our observations are based on a small and not empirically significant sample size. This problem could be addressed by computing the metrics for spreadsheets from other corpuses.

- Our study used a "black box" approach towards the analysis of spreadsheets from our experiment. We just counted reported defects and did not look into their details. Thus, we have no way of determining the amount of false positives among those defects. It is possible that there is a correlation between these Best Practice violations and correctness, once the false positives are filtered out.

- We only analyzed a very small selection of the Best Practices proposed in literature. It is fairly possible that different Best Practices could yield completely different correctness and maintainability indicators than the ones we selected.

- We evaluated Best Practices using a selection of artificially generated spreadsheets from a laboratory experiment which implements a rather simple "which product to buy" scenario. While this might be more realistic than the "wall problem", observation of Best Practices in real-world spreadsheets which solve much more complex tasks might have completely different effects.

- All subjects who took part in the spreadsheet live in the area around Stuttgart. Some of them also have a computer science background. Subjects from other parts of the world with a different background might choose different problem solving approaches for this task.

- Although we exercised many different thresholds for the rules implemented, this does not mean that we exercised all sensible thresholds. There could be some thresholds we missed that would yield a much better correlation than our probes did.

- The tool we used for analysis is a research prototype and not a mature product. While we did not find any indications for faults in our tool when doing basic plausibility checks of its results, we could have overlooked some bugs which might have enormous effects on the computed results.

- We must admit that the final comparison of the results was done with the help of a spreadsheet.

## 7 CONCLUSION

In this study, we tried to find evidence of connections between observance of three common spreadsheet Best Practices and the quality aspects of correctness and maintainability. The results of our experiments with 42 subjects indicate that correctness is not impacted by the observance of the three Best Practices practices inspected. However, initially correct spreadsheets with high observance of Best Practices tend to be the ones whose later modifications yield the most correct results.

Apart from these key findings, this work also makes two secondary contributions to the spreadsheet research community: First, it provides a unique selection of *comparable* spreadsheets which all solve the same tasks but are much more data-intensive than the corpuses from other researchers. Second, it provides a tool which was specially developed for investigating spreadsheet-related research questions. This tool is probably the first open-source spreadsheet analysis tool ever available. It provides an advanced technical framework which other researchers can extend for investigating new research questions in the future.

## ACKNOWLEDGEMENTS

This study would have not been possible without the commitment of all the subjects who took part in our experiments and built the foundation for this study. We are also very grateful to Jochen Ludewig, Kornelia Kuhle and the anonymous reviewers for their constructive criticisms of earlier versions of this paper.